\documentclass[aps,prc,twocolumn,showpacs,showkeys,nofootinbib,superscriptaddress]{revtex4-2}

\usepackage{longtable}
\usepackage{graphicx}
\usepackage{amsmath}
\usepackage{amsfonts}
\usepackage{amssymb}
\usepackage{natbib} 
\usepackage{setspace}
\usepackage{xspace}
\usepackage{cancel}
\usepackage{physics}
\usepackage{tensor,braket,color,bbold,dsfont,slashed}
\usepackage{stackengine} 
\usepackage{mathrsfs}
\usepackage[normalem]{ulem}

\usepackage{hyperref}
\hypersetup{
  colorlinks=true,       
  linkcolor=blue,        
  citecolor=cyan,        
  urlcolor=cyan          
}

\renewcommand{\sout}{\bgroup \color{red} \ULdepth=-.5ex \ULset}

\def\emdash{\nobreak\hspace{0.1em}\textemdash\nobreak\hspace{0.1em}}

\begin{document}
\title{Impact of Charge Symmetry Breaking on Gluon and Sea Quark Distributions in the Pion and Kaon}

\author{Parada T. P. Hutauruk}
\email[Email:~]{phutauruk@gmail.com; phutauruk@pknu.ac.kr}
\affiliation{Department of Physics, Pukyong National University (PKNU), Busan 48513, Korea}
\affiliation{Department of Physics Education, Daegu University, Gyeongsan 38453, Korea}
\date{\today}

\begin{abstract}
In this exploratory study, I present, for the first time, the implications of the charge symmetry breaking (CSB) that arise from the $u$ and $d$ quark mass differences on gluon and sea quark distribution functions of the pion and kaon in the framework of the Nambu--Jona-Lasino (NJL) model, which is a quark-level chiral effective theory of QCD, with the help of the proper-time regularization scheme to simulate color confinement of QCD. From the analysis, one finds that the charge symmetry (CS) gluon distribution for the pion has a good agreement with the prediction results obtained from the recent lattice QCD simulation and JAM global fit QCD analysis at a higher scale of $Q^2 =$ 5 GeV$^2$. The size of the CSB effects on gluon and sea quark distributions for the pion with the realistic ratios of $m_u/m_d = 0.5$ at $Q^2 =$ 5 GeV$^2$ are respectively estimated by 1.3\% and 2.0\% at $x \simeq 1$ in comparison with those for $m_u/m_d = 1.0$, while those for the kaon are approximately about 0.3\% and 0.5\% at $ x \simeq 1$, respectively. A remarkable result is found that the CSB effects on gluon distribution for the kaon are smaller than that for the pion, which has a similar prediction result as that for the CS case.        
\end{abstract}


\maketitle

\section{Introduction} \label{intro}

In the standard model (SM) of quantum chromodynamics (QCD), it is commonly considered that the up and down quarks are identical ($m_u = m_d$, where $m_u$ and $m_d$ are respectively the current masses of the up and down quarks), which is known as a charge symmetry (CS)~\cite{Londergan:2009kj,Miller:2006tv,Miller:1990iz,Slaus:1990nn}. However, empirical evidence~\cite{Londergan:2009kj,Slaus:1990nn,ParticleDataGroup:1994kdp,Gasser:1982ap,Boros:1998qh,Boros:1998es,Boros:1999fy} indicates that the current masses of the up and down quarks are different ($m_u \neq m_d$) that break the CS at a fundamental level of nuclear forces, which is the so-called as the charge symmetry breaking (CSB)~\cite{Glashow:1967rx}. In addition to the CSB, due to the quark-mass difference ($\delta m = m_d - m_u$) between the up and down quarks, it can also be generated by the electromagnetic charge difference that arose from the difference between the electric charge and magnetic moments for the up and down quarks. The electromagnetic-charge difference effects on the quark masses, in nature, are expected to be smaller, which is about one percent level, in comparison with the CSB that emerged from the quark mass difference. Consequently, it makes it more difficult to measure the CSB in experiments or to simulate it in lattice QCD~\cite{Horsley:2015vla}. In the present work, I will focus on the CSB effects arising from the quark mass difference within the pseudoscalar mesons, in particular in the pion ($u\bar{d}$) and kaon ($u\bar{s}$).

Several theoretical and phenomenological studies on the CSB effects arising from the quark mass difference have been observed in the parton distribution functions (PDFs) for the nucleon~\cite{Boros:1998qh,Boros:1998es,Boros:1999fy}, in the electromagnetic form factors (EMFFs), the valence quark distribution functions (VQDFs) of the pion and kaon~\cite{Miller:1990iz,Hutauruk:2018zfk,Hutauruk:2019jja}, and in the hadronic reactions~\cite{Miller:2006tv}. However, mostly, past studies on the CSB have been done so far for the nucleon structure functions. Interestingly, in the CSB~\footnote{Note that, in the literature, the CSB is also a well-known charge symmetry violation (CSV).} nucleon structure analysis of Ref.~\cite{Boros:1998qh}, the authors reported that the CSB effect in the nucleon sea quark distributions is large at small-$x$. With this finding result, analogously, it is expected that at small-$x$, the CSB may have a significant effect on the valence, gluon, and sea quark distribution functions (DFs) for the pion and kaon.

More recently, the CSB effect on the properties for the pion and kaon, EMFFs, and their DFs were studied using the NJL model with the proper-time regularization (PTR) scheme~\cite{Hutauruk:2018zfk,Hutauruk:2019jja}. They found that, for the realistic value of $m_u /m_d \simeq$ 0.5, the CSB has a significant effect on the light-quark sector EMFFs of the pion and kaon. Also, a more interesting result was found that, in the $K^+$ and $K^0$ light-quark sector form factors, the size of the CSB effect is about twice as large in comparison with the up-and down-quark sector EMFFs for the pion. But, in the papers of Refs.~\cite{Hutauruk:2018zfk,Hutauruk:2019jja}, they just focused on the EMFFs and quark distributions for the pion and kaon, they did not yet study the CSB effects on gluon and sea quark distributions for both the pion and kaon. It strongly motivates this present work to investigate the CSB effects on the gluon and sea quark distributions for the pion and kaon, which may provide additional and important information for the future experiments of the Electron-Ion Collider (EIC)~\cite{Arrington:2021biu,Aguilar:2019teb}, the Electron-Ion Collider in China (EicC)~\cite{Anderle:2021wcy}, the COMPASS ++/AMBER new QCD facility at CERN-SPS~\cite{Adams:2018pwt}, and the Solenoidal Large Intensity Device (SoLiD) at JLab12 upgrade and beyond~\cite{zein2022}.

Moreover, besides the impressive progress of the theoretical and phenomenological studies as well as the challenging possibilities for the future experiments mentioned above, a study on the CSB effect becomes more exciting and challenging due to, in fact, other experiments have also been planned and designed at Jefferson Laboratory (JLab) to specifically measure the effects of the CSB on PDFs or other observables \textit{via} the parity-violating deep inelastic scattering (PVDIS) on deuteron~\cite{Miller:2013nea} and pion production in semi-inclusive deep inelastic scattering (SIDIS)~\cite{Londergan:1996vf}. Furthermore, the complementary experiments through the pion-induced Drell Yan reactions~\cite{Londergan:1994gr} and the charged current reactions~\cite{Boer:2011fh} are also very promising tools to provide information about the CSB effects on gluon and sea quark DFs for both the pion and kaon.

In this paper, I investigate, for the first time, the CSB effects arising from the $u$ and $d$ quark mass difference on gluon and sea quark distributions for the pion and kaon using the NJL model with the PTR scheme~\cite{Schwinger:1951nm} that simulates the color confinement of QCD~\cite{Ebert:1996vx}. This exploratory study is expected to give an estimation of the size of CSB effects on partonic structures for the pion and kaon, which can also potentially provide crucial information for extracting the CSB effect from the experiments. The NJL model with the PTR scheme has been widely and successfully applied in the various physics topics of the low-energy nonperturbative QCD such as the VQDFs~\cite{Hutauruk:2016sug}, the fragmentation functions (FFs)~\cite{Matevosyan:2013aka}, the transverse-momentum dependents (TMDs)~\cite{Ninomiya:2017ggn}, the properties for the pion and kaon as well as EMFFs in medium~\cite{Hutauruk:2018qku,Hutauruk:2019was}, the neutron star properties~\cite{Tanimoto:2019tsl}, and the gluon distributions in the nuclear medium~\cite{Hutauruk:2021kej}. It is worth noting that, in the NJL model, there are no gluons and sea quarks dynamics at the initial model scale of $Q_0^2$, since they are already absorbed into the coupling constant of $G_\pi$ in the NJL effective Lagrangian. In the present work, following the works of Refs.~\cite{Hutauruk:2021kej,Wang:2021elw}, the gluon and sea quark distributions for the pion and kaon at a higher factorization scale of $Q^2$ are purely and dynamically generated through the next-leading order (NLO) Dokshitzer-Gribov-Lipatov-Altarelli-Parisi (DGLAP) QCD evolution~\cite{Miyama:1995bd}. Next, the valence quark distribution for the pion is compared to the experimental data (E615)~\cite{Conway:1989fs} and reanalysis E615 data which is so-called E615-M~\cite{Aicher:2010cb}. In the reanalysis data, they included the next-to-leading logarithmic threshold resummation effects in the cross-section of the Drell-Yan process. One is found that the result for the pion valence quark distribution of this work is in excellent agreement with the E615 data~\cite{Conway:1989fs}.

The outline of this paper is organized as follows. In Sec.~\ref{sec:NJLCSB}, I briefly review the NJL effective Lagrangian for the three-flavor NJL model with CSB. I then determine the dressed (constituent) quark mass, meson-quark coupling constant, and meson masses for various values of $m_u/m_d$, which are needed in the calculation of the CSB PDFs for the pion and kaon. In Sec.~\ref{sec:vacpdfcsb}, I present the general expression for the twist-2 DFs for the pion and kaon with CSB. In Sec.~\ref{results}, the numerical results are presented and their implications are discussed. Section~\ref{summary} is devoted to a summary.
%

\section{The Nambu--Jona-Lasinio Model with CSB}
\label{sec:NJLCSB}

In this section, I briefly present the description of the NJL effective Lagrangian and the properties for both the pion and kaon in the NJL model, which is a chiral quark effective theory of QCD. The NJL model maintains the important features of the global symmetry of the nonperturbative QCD such as the spontaneous chiral symmetry breaking (SCSB). The three-flavor NJL Lagrangian can be written in terms of the local four-fermion interaction
\begin{eqnarray}
  \label{eq1pdf}
  \mathscr{L}_{\textrm{NJL}} &=& \bar{q} \left( i \partial \!\!\!/ - \hat{m}_q \right) q + G_\pi [ (\bar{q} \mathbf{\lambda}_a q)^2 -(\bar{q} \mathbf{\lambda}_a \gamma_5 q )^2 ] \nonumber \\
  &-& G_\rho [(\bar{q} \mathbf{\lambda}_a \gamma^\mu)^2 + (\bar{q} \mathbf{\lambda}_a \gamma^\mu \gamma_5 q)^2].
\end{eqnarray}
With the quark fields $q$ are defined by $q = (u, d, s)^T$, $\hat{m}_q = \textrm{diag}(m_u, m_d, m_s)$ represents the current-quark mass matrix, and $\mathbf{\lambda}_a$ are the Gell-Mann matrices in flavor space with $\lambda_0 \equiv \sqrt{\frac{2}{3}} \mathds{1}$. The $G_\pi$ and $G_\rho$ are the local four-fermion coupling constants, which have dimensional units of GeV$^{-2}$. Thus, a standard solution to the NJL gap equation is given by
\begin{eqnarray}
  \label{eq2pdf}
  S_q^{-1} (p) &=& p\!\!\!/ - M_q + i \epsilon, 
\end{eqnarray}
where the dressed quark propagator is defined by $S_q (p) = \mathrm{diag}[S_u (p), S_d (p), S_s (p)]$ in the quark flavor space. The dynamical (constituent) quark mass in the PTR scheme is given by
\begin{eqnarray}
  \label{eq3pdf}
  M_q &=& m_q + \frac{3G_\pi M_q}{\pi^2} \int_{\tau_{\textrm{UV}}^2}^{\tau_{\textrm{IR}}^2} \frac{d\tau}{\tau^2} \exp \left[ -\tau M_q^2 \right],
\end{eqnarray}
where $\tau_{\textrm{IR}}^2 = 1/\Lambda_{\textrm{IR}}^2$ and $\tau_{\textrm{UV}}^2 = 1 /\Lambda_{\textrm{UV}}^2$ stand for the infrared (IR) and ultraviolet (UV) integration limits with the value of $\Lambda_{\textrm{IR}} =$ 0.240 GeV, which is determined based on the limit of $\Lambda_{\textrm{QCD}}$ ($\simeq$ 0.2-0.3 GeV), and $\Lambda_{\textrm{UV}}$ are the IR and UR PTR scheme cutoffs, respectively. In addition. the $\Lambda_{\textrm{UV}}$ is applied to remove the poles at $\tau = 0$ to have a finite theory, whereas the $\Lambda_{\textrm{IR}}$ is used to remove the particle propagation for larger value of $\tau$~\cite{Ebert:1996vx}.

In the NJL model, pions and kaons as the relativistic bound state of the dressed quark-antiquark can be evaluated by solving the Bethe-Salpeter equations (BSEs). The solutions for the BSEs are given by the corresponding interaction channels of the two-body amplitude. For those pions and kaons, it is given by
\begin{eqnarray}
  \label{eq4pdf}
  t_{[\mathrm{\pi}, \mathrm{K}]} (p) &=& \frac{-2i G_\pi}{1 + 2 G_\pi \Pi_{[\mathrm{\pi}, \mathrm{K}]} (p^2)},
\end{eqnarray}
where the bubble diagrams or polarization insertions for the pion and kaon are respectively given by
\begin{eqnarray}
  \label{eq5pdf}
  \Pi_{\mathrm{\pi}^+} (p^2) &=& 6i \int \frac{d^4 k}{ (2 \pi)^4} [\gamma_5 S_u (k) \gamma_5 S_{\bar{d}} (k+p)], \\
  \label{eq5pdfb}
  \Pi_{\mathrm{K}^+} (p^2) &=& 6i \int \frac{d^4 k}{ (2 \pi)^4} [\gamma_5 S_u (k) \gamma_5 S_{\bar{s}} (k+p)].
\end{eqnarray}
Here, the subscripts of $u$, $\bar{d}$, and $\bar{s}$ are, respectively, the up, anti-down, and anti-strange quarks. From the pole of the amplitude $t_{[\mathrm{\pi},\mathrm{K}]} (p)$ of Eq.~(\ref{eq4pdf}), I determine the pion and kaon masses by solving the pole equations $1 + 2 G_\pi \Pi_\pi (p^2 = m_\pi^2)=$ 0 for the pion and $1 + 2 G_\pi \Pi_K (p^2 = m_K^2)=$ 0 for the kaon. By solving these equations analytically, the expressions for the pion and kaon masses in the PTR scheme are obtained by
\begin{eqnarray}
  \label{eq6pdf}
  m_{\mathrm{\pi}^2} &=&  \left( \frac{m_u}{M_u} + \frac{m_{\bar{d}}}{M_{\bar{d}}} \right) \frac{1}{G_\pi \mathcal{I}_{u\bar{d}} (m_\pi^2)} + (M_u - M_{\bar{d}})^2 , \\
   \label{eq6pdfb}
  m_{\mathrm{K}^2} &=& \left( \frac{m_u}{M_u} + \frac{m_{\bar{s}}}{M_{\bar{s}}} \right) \frac{1}{G_\pi \mathcal{I}_{u \bar{s}} (m_K^2)} + (M_{\bar{s}} - M_u)^2 ,
\end{eqnarray}
where the quantity of $\mathcal{I}_{u\bar{s}} (p^2)$ in Eq.~(\ref{eq6pdfb}) is defined for the kaon by
\begin{eqnarray}
  \label{eq7pdf}
  \mathcal{I}_{u\bar{s}} (p^2) &=& \frac{3}{\pi^2} \int_0^1 dz \int_{\tau_{\textrm{UV}^2}}^{\tau_{\textrm{IR}^2}} \frac{d\tau}{\tau} \nonumber \\
  &\times& \exp \left( -\tau (z (z-1) p^2 + z M_{\bar{s}}^2 + (1-z) M_u^2) \right).
\end{eqnarray}
Analogously, for the pion case, one of the quark flavors in Eq.~(\ref{eq6pdf}) is replaced by $\bar{s} \to \bar{d}$, resulting $\mathcal{I}_{u \bar{d}} (p^2) $. The dressed quark masses that appear in Eqs.~(\ref{eq6pdf}) and~(\ref{eq6pdfb}) indicate the nature of the Goldstone boson of the pion and kaon.

The meson-quark coupling constants for the pion and kaon mesons can be easily evaluated \textit{via} the residue at a pole in the quark-antiquark amplitude $t_{[\mathrm{\pi},\mathrm{K}]} (p)$, which can be simply determined from the first derivative of the polarization insertions (bubble diagrams) in Eqs.~(\ref{eq5pdf}) and~(\ref{eq5pdfb}) with respect to the $p^2$ and one has
\begin{eqnarray}
  \label{eq8pdf}
  Z_{[\mathrm{\pi}, \mathrm{K}]}^{-1} = g_{m q\bar{q}} ^{-2} &=& -\frac{\partial \Pi_m (p^2)}{\partial p^2} \Big|_{p^2 = m_m^2}, 
\end{eqnarray}
with the subscript of $m = [\pi, K]$. Note that the meson-quark coupling constants are related to the wave-function renormalization constant, which gives $Z_{[\mathrm{\pi}, \mathrm{K}]} = g_{[\mathrm{\pi}, \mathrm{K}]q\bar{q}} ^{-2}$. Here, I am interested in the effects of the CSB on gluon and sea quark distributions for the pion and kaon. Next, further, detailed explanations of the quark distributions for the pion and kaon and the DGLAP QCD evolution, where the gluon and sea quark distributions are obtained, will be described in Sec.~\ref{sec:vacpdfcsb}.

\section{The CSB on Pion and kaon parton distributions}
\label{sec:vacpdfcsb}
%

In this section, I present the equations for the leading-twist quark distributions for the pion and kaon. The twist-2 quark distributions for both the pion and kaon are simply given by
\begin{eqnarray}
  \label{eq9pdf}
  q_{[\pi,K]} (x) &=& \frac{p^{+}}{2\pi} \int d\xi^{-} \exp \left( ix p^{+} \xi^{-} \right) \langle [\pi,K]|J^+|[\pi,K] \rangle_{c}, \nonumber \\
\end{eqnarray}
where $J^+=\bar{q}(0) \gamma^{+} q (\xi^{-})$ and $x = \frac{k^{+}}{p^{+}}$ is the Bjorken scaling variable or the longitudinal momentum fraction of the parton for both the pion and kaon mesons with $k^{+}$ is the plus-component of the struck momentum of quark and $p^{+}$ is the plus-component of the pion and kaon meson (parent hadron) momentum, $\xi$ stands for the variable of the skewness, and the subscript $c$ denotes the connected matrix element. 
%
\begin{figure}[b]
  \centering\includegraphics[width=0.85\columnwidth]{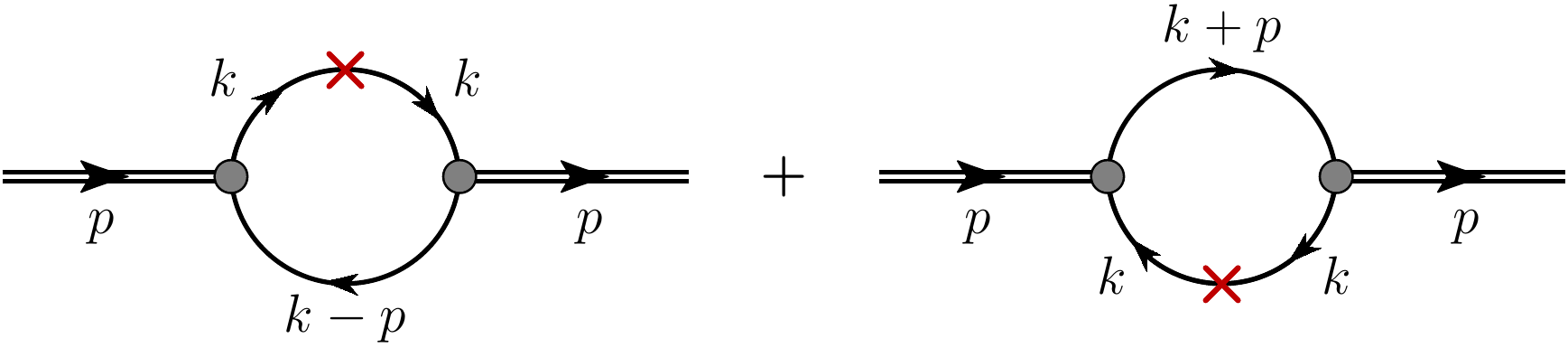}
  \caption{\label{fig1a} The relevant Feynman diagrams for both the pion and kaon valence-quark DFs. The red cross represents an operator insertion of 
    $\gamma^+ \delta \left( p^+x - k^+ \right) \hat{P}_q$, where $\hat{P}_q$ is the projection operator for quarks of flavor $q$.}
\end{figure}

Following the previous works of Refs.~\cite{Hutauruk:2016sug,Hutauruk:2021kej}, one can then evaluate the quark distributions for both the pion and kaon based on two relevant Feynman diagrams in Fig.~\ref{fig1a}. The equations of the operator insertion for the light and strange quarks are defined by
\begin{eqnarray}
  \label{eq10pdf2}
\gamma^+ \delta \left( k^{+}-xp^{+} \right) \hat{P}_{l} &=& \gamma^+ \delta \left( k^{+} - x p^{+} \right) \frac{1}{2} \left[ \frac{2}{3} \mathds{1} \pm \lambda_3 + \frac{1}{\sqrt{3}} \lambda_8 \right],
\cr
\gamma^+ \delta \left( k^{+}-xp^{+} \right) \hat{P}_{s} &=& \gamma^+ \delta \left( k^{+} - x p^{+} \right) \left[ \frac{1}{3} \mathds{1} - \frac{1}{\sqrt{3}} \lambda_8 \right].
\end{eqnarray}
Taking the relation $\bar{q} (x) = - q(-x)$ into account, the valence quark and anti-quark distributions for both the pion and kaon are respectively given by
\begin{eqnarray}
  \label{eq10pdf}
  q_{[\pi,K]} (x) &=& i g_{m q \bar{q}}^2 \int \frac{d^4k}{(2\pi)^4} \delta \left( k^{+} - xp^{+} \right) \nonumber \\
  &\times& \mathrm{Tr}_{c,f,\gamma} [\gamma_5 \lambda_a^{\dagger} S_l (k) \gamma^{+} \hat{P}_{u/d} S_l (k) \gamma_5 \lambda_a S_l (k-p)], \nonumber \\
  \bar{q}_{[\pi,K]} (x) &=& -i g_{m q \bar{q}}^2 \int \frac{d^4k}{(2\pi)^4} \delta \left( k^{+} + xp^{+} \right) \nonumber \\
  &\times& \mathrm{Tr}_{c,f,\gamma} [\gamma_5 \lambda_a S_l (k) \gamma^{+} \hat{P}_{\bar{d}/\bar{s}} S_l (k) \gamma_5 \lambda_a^{\dagger} S_s (k+p)], \nonumber \\
\end{eqnarray}
where the meson-quark coupling constant is similarly defined in Eq.~(\ref{eq8pdf}). The trace runs over the color, flavor, and Lorentz indices. Using similar procedures as in Refs.~\cite{Hutauruk:2016sug,Hutauruk:2021kej}, the valence quark distributions for both the pion and kaon are straightforwardly evaluated \textit{via} the moment, which can be defined by
\begin{eqnarray}
  \label{eq11pdf}
  \mathscr{A}_n &=& \int_0^1 dx x^{[n-1]} q_{[\mathrm{\pi},\mathrm{K}]} (x),
\end{eqnarray}
where $n$ is an integer number, which is used to remove the $\delta-$function in Eq.~(\ref{eq10pdf}). Next, one performs the Ward-Takahashi-like identity (WTI) using $S(k) \gamma^+ S(k) = - \partial S(k) / \partial k_{+}$, the Feynman parametrization, and the PTR scheme. Finally, I then obtain the CSB valence quark DFs for the $\pi^+$ in the PTR scheme, which gives
\begin{eqnarray}
  \label{eq12pdf}
  q_{{\pi}^+} (x) &=& \frac{3 g_{\pi q \bar{q}}^2}{4\pi^2} \int_0^1 dx \int^{\tau_{IR}^2}_{\tau_{UV}^2} d\tau \nonumber \\
  &\times& \exp \left( -\tau (x(x-1)m_\pi^2 + xM_{\bar{d}}^2 + (1-x) M_u^2) \right) \nonumber \\
  &\times& \Big[ \frac{1}{\tau} + x (1-x) \left( m_\pi^2 - (M_u -M_{\bar{d}})^2 \right)\Big],
\end{eqnarray}
and for the CSB valence antiquark distribution for the $\pi^+$ is given by
\begin{eqnarray}
  \label{eq12pdfb}
  \bar{q}_{{\pi}^+} (x) &=& \frac{3 g_{\pi q \bar{q}}^2}{4\pi^2}  \int_0^1 dx \int^{\tau_{IR}^2}_{\tau_{UV}^2} d\tau \nonumber \\
  &\times& \exp \left( -\tau (x(x-1)m_\pi^2 + xM_u^2 + (1-x) M_{\bar{d}}^2) \right) \nonumber \\
  &\times& \Big[ \frac{1}{\tau} + x (1-x) \left( m_\pi^2 - (M_u -M_{\bar{d}})^2 \right)\Big].
\end{eqnarray}
Analogously, for the kaon case, the final expression for the CSB valence quark DFs can be obtained by replacing $M_{\bar{d}} \rightarrow M_{\bar{s}}$ and $g_{\pi q \bar{q}}^2 \rightarrow g_{K q \bar{q}}^2$ in Eqs.~(\ref{eq12pdf}) and~(\ref{eq12pdfb}).

The valence quark DFs with CSB for both the pion and kaon must preserve the baryon number and momentum sum rules. The baryon number and momentum conservations for both the pion and kaon can be expressed by
\begin{eqnarray}
  \label{eq13pdf}
  \int_0^1 dx [u_K (x) - \bar{u}_K (x)] = \int_0^1 dx [\bar{s}_K -s_K (x)] &=& 1, \nonumber \\
  \int_0^1 dx\,x[u_K (x) + \bar{u}_K (x) + s_K (x) + \bar{s}_K (x)] &=& 1. 
\end{eqnarray}
Here, reminding the reader again that, in the NJL model, the sea quarks and gluons distributions for both the pion and kaon mesons are zero at an initial NJL model scale of $Q_0^2$, since in the NJL model, the gluons are integrated out from the NJL Lagrangian and absorbed into the $G_\pi$ coupling constant. Therefore, the NJL model has no gluons and sea quarks dynamics that are adopted from Refs.~\cite{Wang:2021elw,Hutauruk:2021kej}. Hence, in the present work, both gluon and sea quark distributions are purely and dynamically generated using the NLO DGLAP QCD evolution. It now turns to explain the QCD evolution, where the parton distributions for both the pion and kaon are obtained. The valence quark distribution, which is the so-called non-singlet (NS) quark distribution, is defined by
\begin{eqnarray}
  q_{\mathrm{NS}} (x) &=& q(x) - \bar{q} (x),
\end{eqnarray}
where $q (x)$ and $\bar{q}(x)$ are respectively the quark and antiquark distributions. The NS quark distributions in the QCD evolution via DGLAP are defined by
\begin{eqnarray}
  \frac{\partial q_{\textrm{NS}} (x,Q^2)}{\partial \ln (Q^2)} &=& \mathbf{P}_{qq} \left(x,\alpha_{\textrm{s}}(Q^2)\right) \otimes q_{\textrm{NS}} (x,Q^2),
\end{eqnarray}
where $\mathbf{P}_{qq}$ stands for the splitting function of the quark (q)-quark (q). $\mathbf{P}_{qq}$ is physically interpreted as a probability for a quark of type $q$ with momentum fraction $z$ emitting the quark and becomes a new type of quark $q$ with momentum fraction $x$. The product convolution between the splitting function and the non-singlet quark distribution is given by
\begin{eqnarray}
  \mathbf{P}_{qq} \otimes q_{\mathrm{NS}} &=& \int_x^1 \frac{dz}{x} \mathbf{P}\left(\frac{x}{z} \right) q_{\mathrm{NS}}(z,Q^2).
\end{eqnarray}
Another type of quark distribution is the so-called singlet quark distribution and it can be expressed by
\begin{eqnarray}
  q_\mathrm{S} (x) = \sum_i q_i^+ = \sum_i q_i (x) + \bar{q}_i (x),
\end{eqnarray}
where $i$ is the quark flavor. The singlet quark distributions are given by
\begin{eqnarray}
  \label{eqdglap}
  \frac{\partial}{\partial \ln Q^2} \begin{bmatrix} q_\mathrm{S} (x,Q^2) \\ g(x,Q^2) \end{bmatrix} &=& \begin{bmatrix} \mathbf{P}_{qq} & \mathbf{P}_{qg} \\ \mathbf{P}_{gq} & \mathbf{P}_{gg} \end{bmatrix} \otimes \begin{bmatrix} q_\mathrm{S} (x,Q^2) \\ g(x,Q^2) \end{bmatrix}.
\end{eqnarray}
The gluon distributions for the pion and kaon can be numerically determined by solving Eq.~(\ref{eqdglap}). Analogously, following the Taylor expansion series, the splitting functions can be expanded in terms of the running coupling constant $\alpha_{s} (Q^2)$ in the perturbation region and one has
\begin{eqnarray}
  \mathbf{P} (z,Q^2) &=& \left[\frac{\alpha}{2\pi} \right] \mathbf{P}^{(0)} (z) + \left[\frac{\alpha}{2\pi} \right]^2 \mathbf{P}^{(1)} (z) + \cdot \cdot \cdot,
\end{eqnarray}
where the $\mathbf{P}^{(0)} (z)$ stands for the leading order (LO), and the second term in $\mathbf{P}^{(1)} (z)$ represents the NLO. In the literature, splitting function expression up to next-next leading order (NNLO) for all parton distributions can be found in Ref.~\cite{Cafarella:2005zj,Devee:2012zz}. However, for the consistent solution of the DGLAP equation, in this work, I also need $\alpha_s(Q^2)$ up to the NLO. The NLO result of $\alpha_{s} (Q^2)$ is defined by
\begin{eqnarray}
  \alpha_{s} (Q^2) &=& \frac{4\pi}{\beta_0} \frac{1}{\ln (Q_{\Lambda})} \left[ 1 - \frac{\beta_1}{\beta_0} \frac{\ln \ln (Q_{\Lambda})}{\ln (Q_{\Lambda})}\right] + \mathscr{O} \left( \frac{1}{\ln^2 (Q_{\Lambda})}\right), \nonumber \\
\end{eqnarray}
with
\begin{eqnarray}
  Q_{\Lambda} &=& \frac{Q^2}{\Lambda_{\mathrm{QCD}}^2},~~~~~\beta_0 = \frac{11}{3}N_c - \frac{4}{3}N_f,  \\
  \beta_1 &=& \frac{34}{3} N_c^2 - \frac{10}{3} N_c N_f - 2 C_F N_f,
\end{eqnarray}
where $N_c$ and $N_f$ are the numbers of colors and active flavors, respectively. $ C_F = \frac{4}{3}$ and the value of $\Lambda_{\mathrm{QCD}}$ depends on the numbers of active flavors and renormalization scheme. More details about the DGLAP QCD evolution can be found in Refs.~\cite{Miyama:1995bd,Cafarella:2003jr,Altarelli:1977zs,Gribov:1972ri,Kogut:1973ub}.

\section{Numerical result} \label{results}

\begin{figure*}[t]
  \centering{\includegraphics[width=0.45\textwidth]{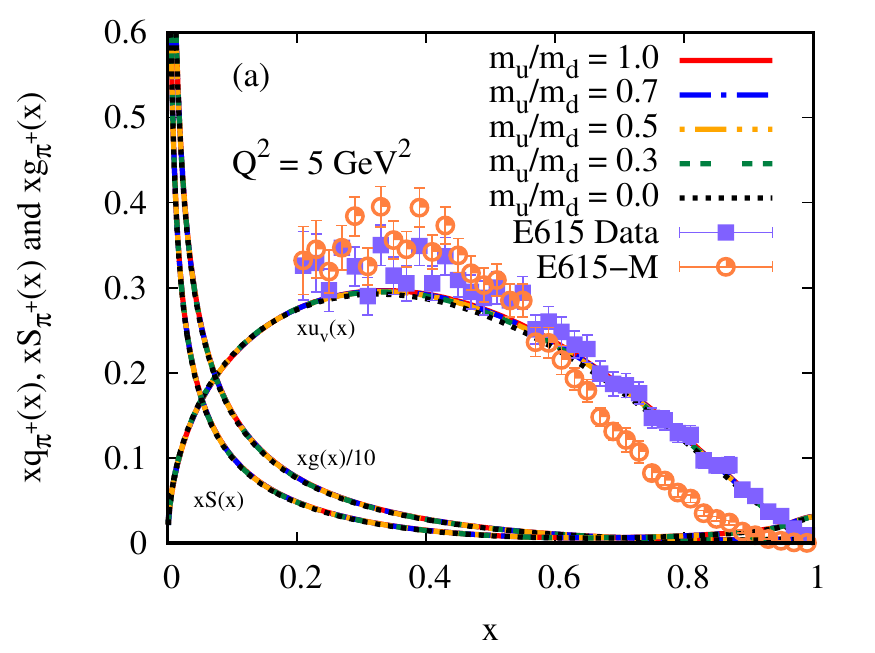}}
  \centering{\includegraphics[width=0.45\textwidth]{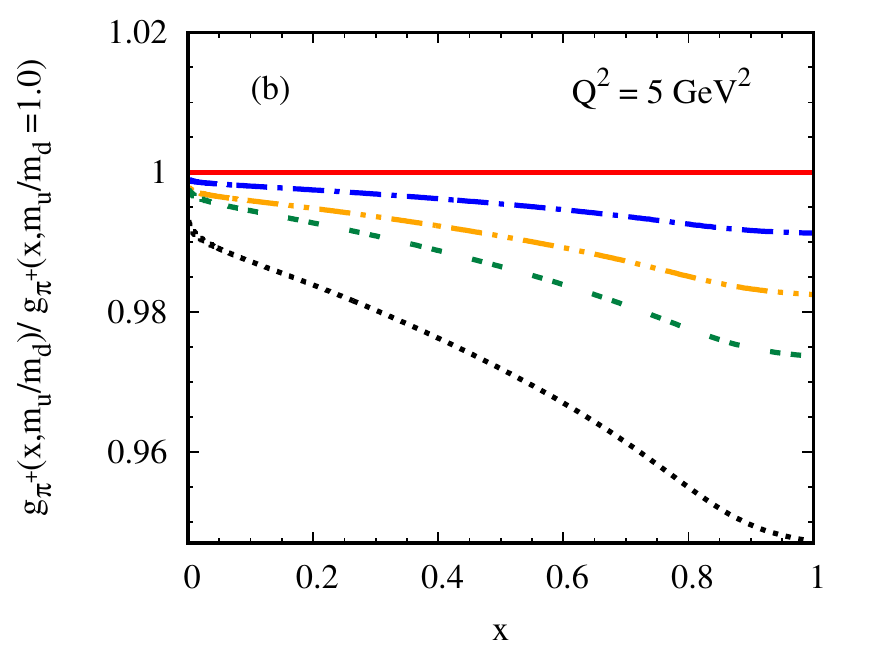}}
  \centering{\includegraphics[width=0.45\textwidth]{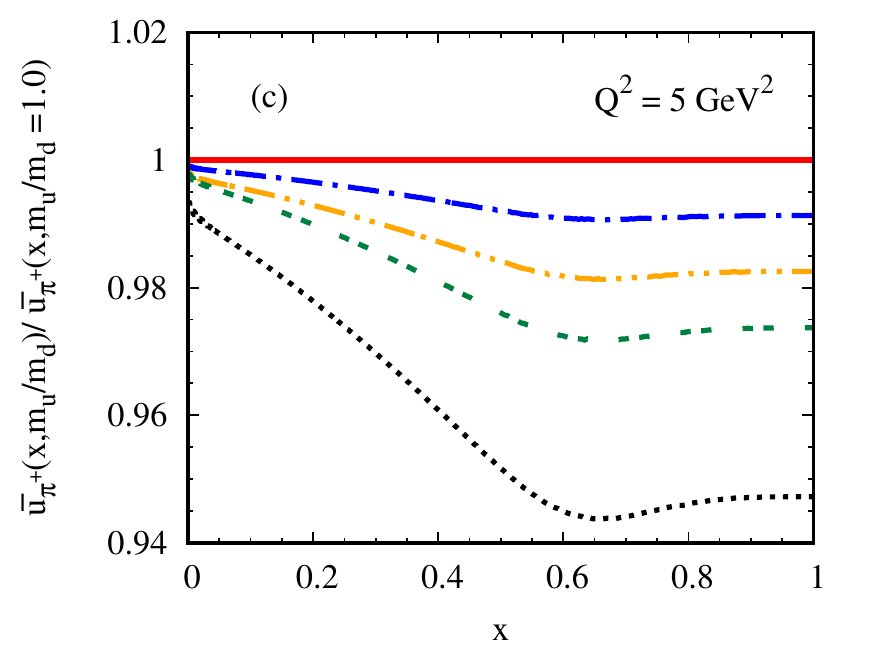}}
  \centering{\includegraphics[width=0.45\textwidth]{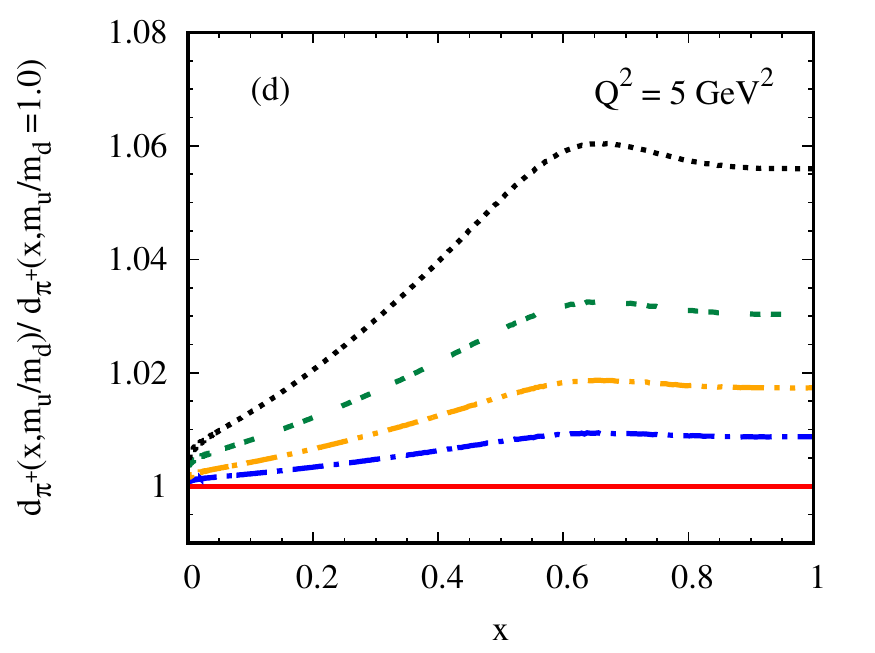}}
  \caption{\label{fig1} (a) valence quark, sea quark, and gluon distributions for the pion as a function of the longitudinal momentum of $x$ with various values of $m_u/m_d$ after evolving via the LO$+$NLO calculations at $Q^2 =$ 5 GeV$^2$. Experimental data are taken from Refs.~\cite{Conway:1989fs,Aicher:2010cb}, (b) ratios of the gluon distributions for the pion with various values of $m_u/m_d$ to those with $m_u/m_d = 1.0$ as a function of the $x$, (c) ratios of the sea up-quark distributions for the pion with various values of $m_u/m_d$ to those with $m_u/m_d = 1.0$ as a function of $x$, and (d) ratios of the sea down-quark distributions for the pion with various values of $m_u/m_d$ to those with $m_u/m_d = 1.0$ as a function of $x$.}
\end{figure*}

\begin{figure*}[t]
  \centering{\includegraphics[width=0.45\textwidth]{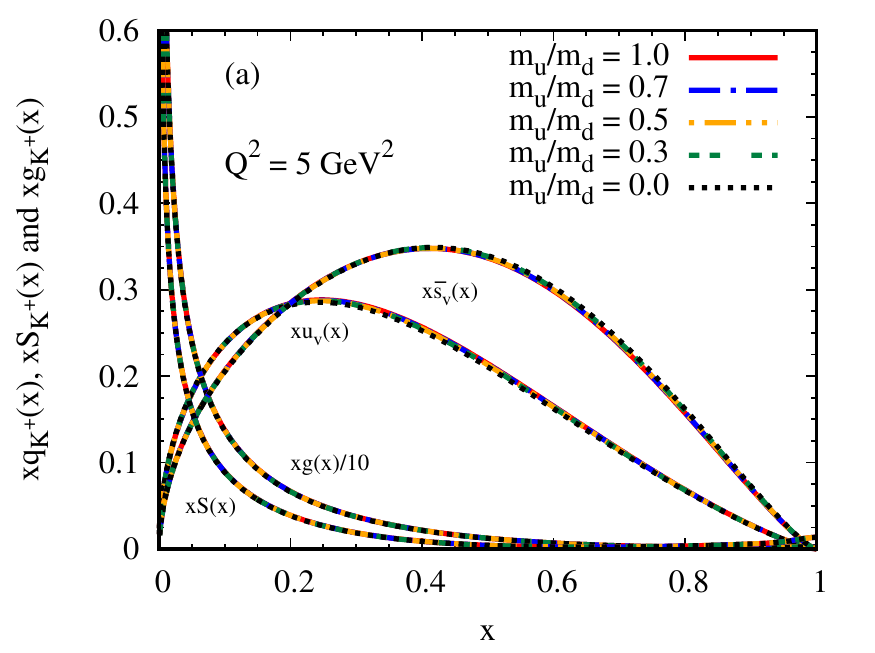}}
  \centering{\includegraphics[width=0.45\textwidth]{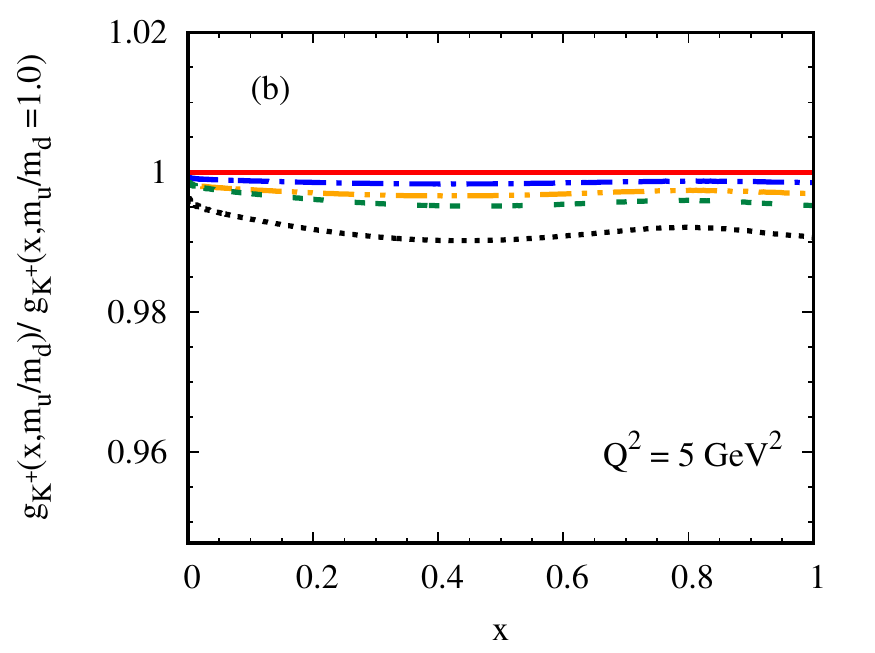}}
  \centering{\includegraphics[width=0.45\textwidth]{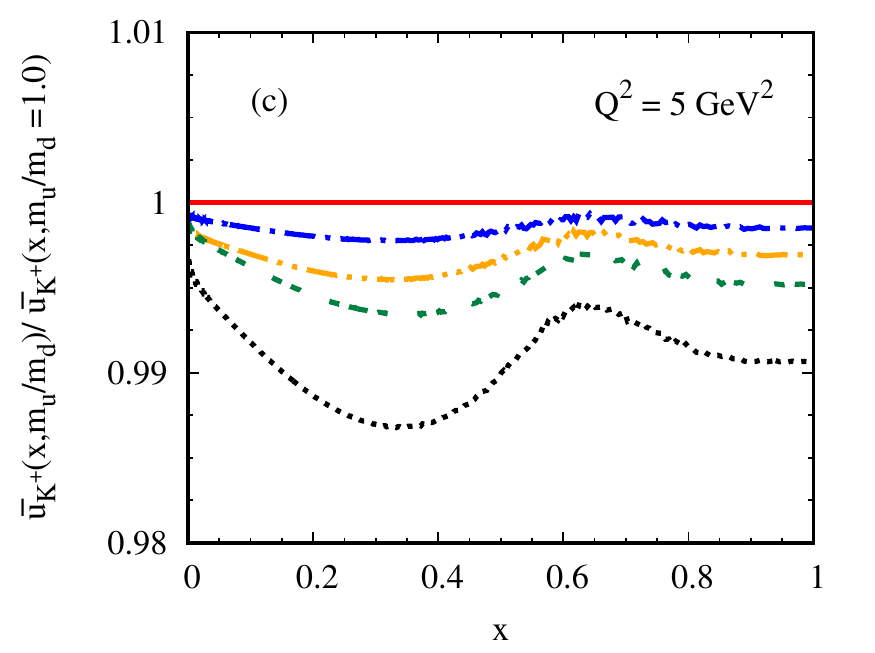}}
  \centering{\includegraphics[width=0.45\textwidth]{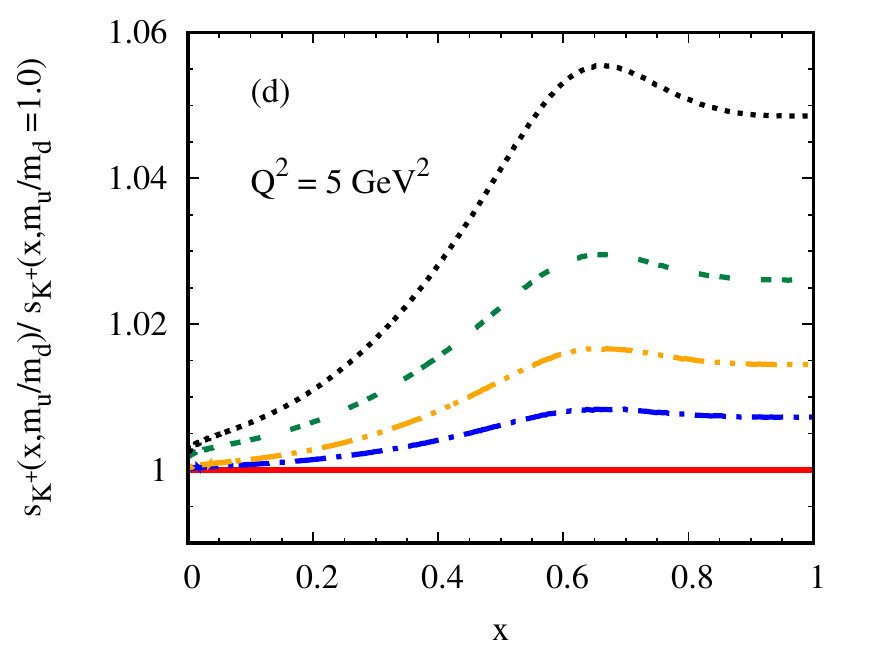}}
  \caption{\label{fig2} (a) valence quark, sea quark, and gluon distributions for the kaon as a function of the longitudinal momentum$x$ with various CSB values of $m_u/m_d$ after evolving via the LO$+$NLO calculations at $Q^2 =$ 5 GeV$^2$, (b) ratios of the gluon distributions for the kaon with various values of $m_u/m_d$ to those with $m_u/m_d = 1.0$ as a function of the $x$, (c) ratios of the sea up-quark distributions for the kaon with various values of $m_u/m_d$ to those with $m_u/m_d = 1.0$ as a function of the $x$, and (d) ratios of the sea $s-$quark distributions for the kaon with various values of $m_u/m_d$ to those with $m_u/m_d = 1.0$ as a function of the $x$.}
\end{figure*}

\begin{figure*}[t]
  \centering{\includegraphics[width=0.45\textwidth]{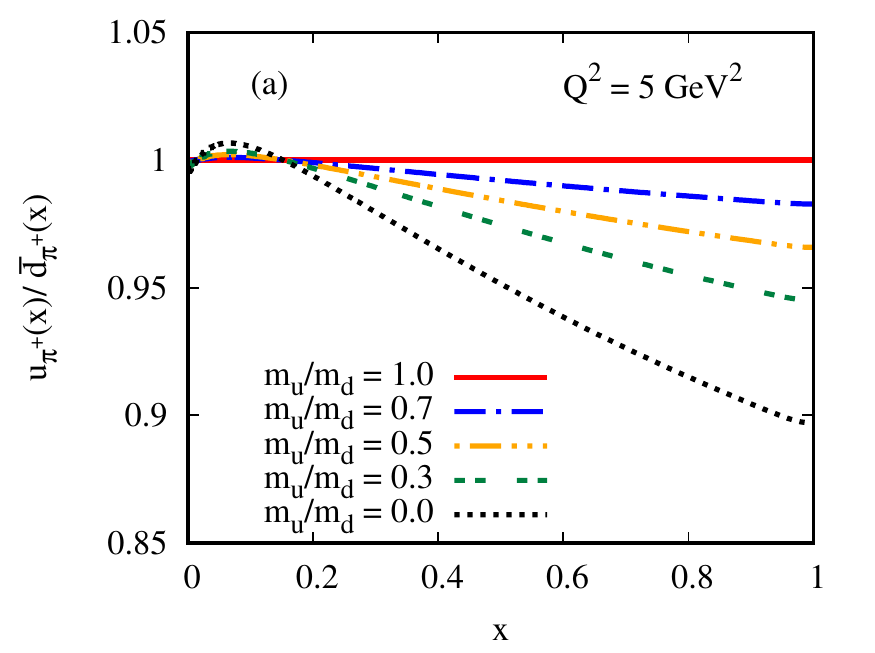}}
  \centering{\includegraphics[width=0.45\textwidth]{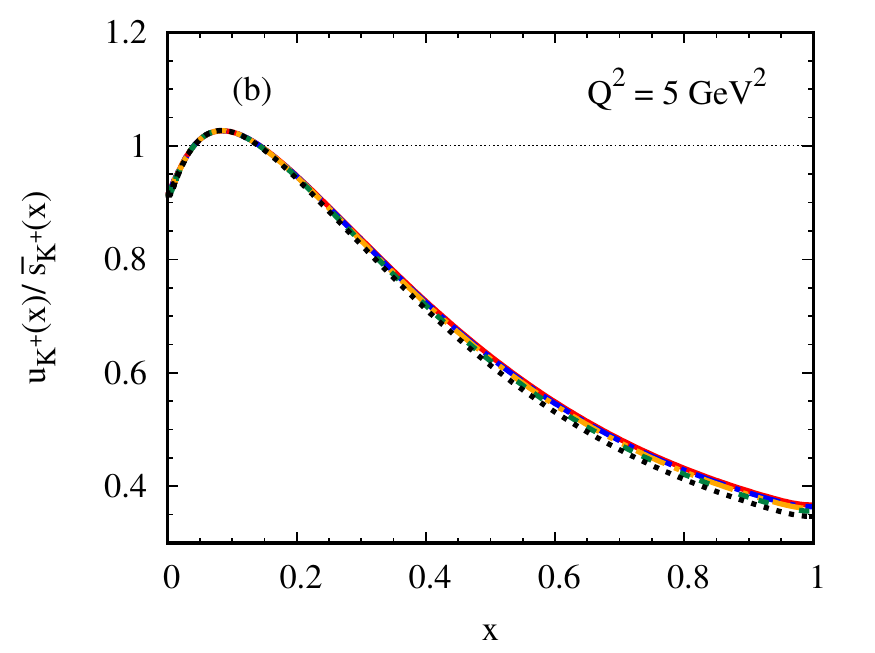}}
  \centering{\includegraphics[width=0.45\textwidth]{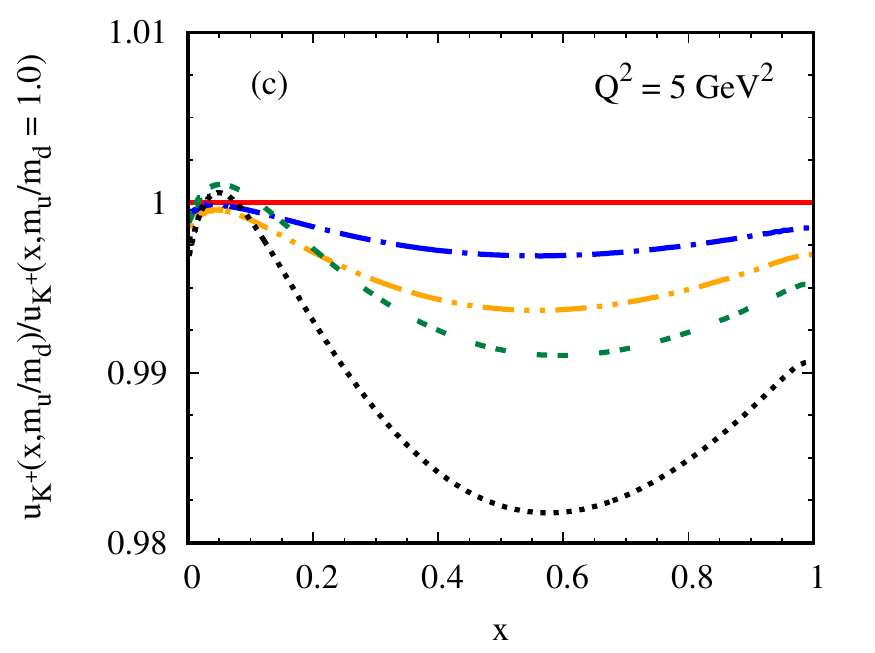}}
  \centering{\includegraphics[width=0.45\textwidth]{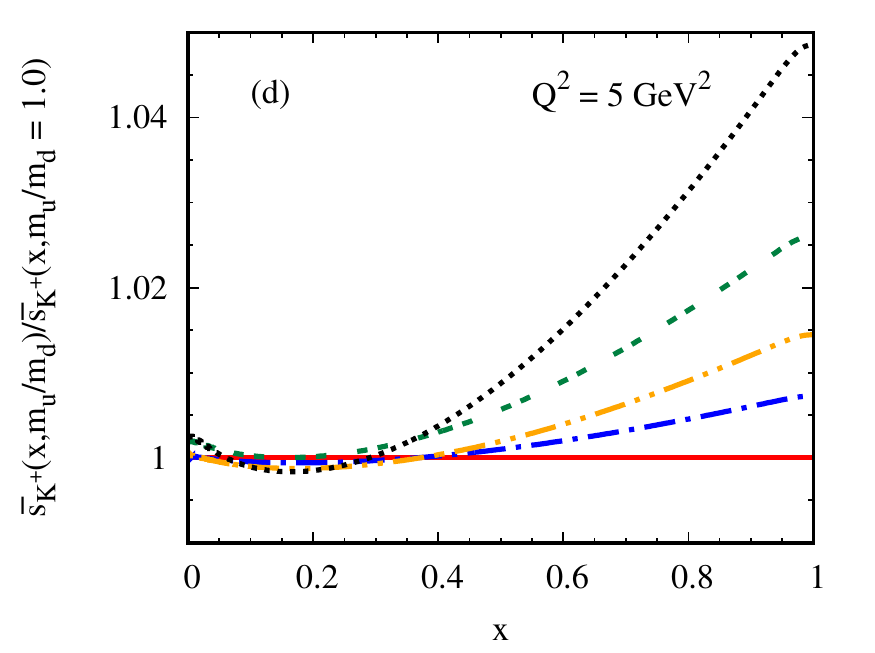}}
  \caption{\label{fig3} (a) Ratios of the up to anti-down quark distributions for the pion as a function of the longitudinal momentum$x$ with various values of $m_u/m_d$ after evolving \textit{via} the LO$+$NLO calculations at $Q^2 =$ 5 GeV$^2$, (b) ratios of the up to anti-strange quark distributions for the kaon with various values of $m_u/m_d$ as a function of the $x$, (c) ratios of the up quark distributions for the kaon with various values of $m_u/m_d$ to those with $m_u/m_d = 1.0$ as a function of the $x$, and (d) ratios of the anti-strange quark distributions for the kaon with various values of $m_u/m_d$ to those with $m_u/m_d = 1.0$ as a function of the $x$.}
\end{figure*}

\begin{figure*}[t]
  \centering{\includegraphics[width=0.45\textwidth]{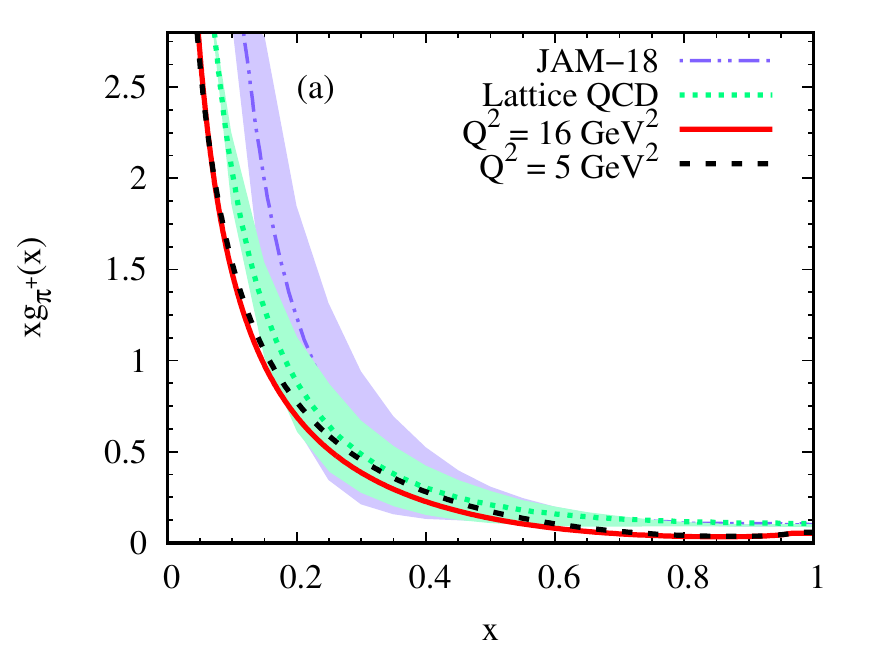}}
  \centering{\includegraphics[width=0.45\textwidth]{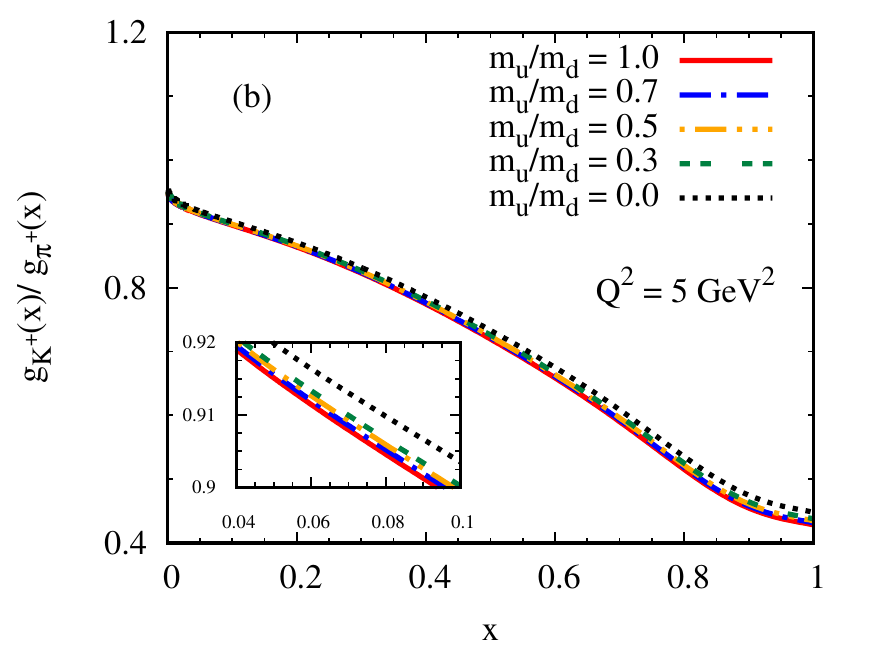}}
  \caption{\label{fig4} (a) Gluon distributions for the pion at $Q^2 =$ 5 GeV$^2$ and $Q^2 =$ 16 GeV$^2$ as a function of $x$ in comparison with the prediction results from the lattice QCD simulation~\cite{Fan:2021bcr} and JAM global QCD fit analysis~\cite{Barry:2018ort} and (b) ratios of the gluon distributions for the kaon to those for the pion with various $m_u/m_d$ at $Q^2 =$ 5 GeV$^2$.}
\end{figure*}

Here, the numerical results for the effects of CSB on the gluon and sea quark distributions for the pion and kaon with various values of $m_u/m_d$ in comparison with the lattice QCD simulation~\cite{Fan:2021bcr} and Jefferson Lab Angular Momentum Collaboration (JAM) phenomenology global fit QCD analysis~\cite{Barry:2018ort} are presented. In the present work, the parameters of the NJL model have to be determined are the current quark masses: $m_u$, $m_d$, and $m_s$, the four-fermion coupling constants: $G_\pi$ and $G_\rho$, and the PTR cutoff parameters: $\Lambda_{\mathrm{UV}}$. For given the values of $m_u/m_d$, I then determine $G_\pi$, $G_\rho$, and $\Lambda_{\mathrm{UV}}$ by fitting the physical values of $m_{\pi^+} =$ 140 MeV, and $f_{\pi^+} =$ 93 MeV. The strange current-quark mass is evaluated to fit the physical kaon mass $m_{K^+} =$ 495 MeV. The value of the PTR IR cutoff is set to $\Lambda_{\mathrm{IR}} =$ 240 MeV which is similar to the order of $\Lambda_{\mathrm{QCD}}$ and $m_\rho = 775 $ MeV. Note that the average current quark mass of $m_0 = \frac{1}{2} \left( m_u + m_d \right)$ is determined to give the $M_0 =$ 400 MeV in the chiral limit. I obtain the $m_s =$ 356 MeV, $M_s =$ 611 MeV, and $m_0 =$ 16.4 MeV. It gives the ratios of $m_s/m_0 =$ 22 which is in good agreement with the empirical value of $\frac{2m_s}{m_u +m_d} = 27.5 \pm 1.0$~\cite{ParticleDataGroup:2012pjm,Hutauruk:2019jja}. The complete results for the CSB effects on the NJL parameters for the dressed quark masses, the pion and kaon masses, the meson-quark coupling constants, and other related quantities for various values of $m_u/m_d$ are shown in Table.~\ref{tab1}. Similar results for the NJL parameters can be found in Ref.~\cite{Hutauruk:2019jja}. Next, these obtained NJL parameters will be used in the calculations of the quark, gluon, and sea quark distributions for the pion and kaon as shown in Figs.~\ref{fig1}-\ref{fig4}.
%
%
\begin{table*}[tbp]
  \caption{\label{tab1} Summary results for the NJL parameters: current-and constituent-quark masses, neutral pion,
    kaon masses, meson leptonic decay constant, meson-quark coupling constants for various values of $m_u/m_d$.
    All quantities are in units of MeV with the exception of $G_{\pi,\rho}$ that are in units of GeV$^{-2}$.}
\addtolength{\tabcolsep}{5pt}
\begin{tabular}{ccccccccccccc}
\hline\hline
$m_u/m_d$ & $m_u$ & $m_d$ & $M_u$ & $M_d$ & $m_{\pi^0}$ & $m_{K^\pm}$ & $f_{\pi^0}$ & $Z_{\pi^\pm}$ & $Z_{K^\pm}$ & $G_\pi$ & $G_\rho$ & $\Lambda_{\rm UV}$ \\[0.2em]
\hline
1.0       & 16.4 & 16.4 & 400 & 400 & 140.00    & 495 & 93.00  & 17.853 & 20.89 & 19.04 & 10.778 & 644.87 \\
0.7      & 13.5 & 19.3 & 398 & 402 & 139.93 & 493 & 92.99 & 17.853 & 20.86 & 19.04 & 10.776 & 644.86 \\
0.5      & 11.0 & 21.9 & 396 & 404 & 139.76 & 491 & 92.98 & 17.852 & 20.83  & 19.05 & 10.773 & 644.83 \\
0.3      & 7.58 & 25.3 & 393 & 406 & 139.38 & 489 & 92.95 & 17.850 & 20.80 & 19.05 & 10.764 & 644.77 \\
0.0       & 0.0    & 32.9 & 387 & 412 & 137.84 & 483 & 92.83 & 17.842 & 20.73 & 19.06 & 10.731 & 644.52 \\
\hline\hline
\end{tabular}
\end{table*}

Results for the valence quark, gluon, and sea quark distributions for the pion as well as the ratios between the quark distributions for the pion with various values of $m_u/m_d$ after evolving \textit{via} the NLO DGLAP QCD evolution from an initial model scale of $Q_0^2 =$ 0.16 GeV$^2$ to a higher factorization scale of $Q^2 =$ 5 GeV$^2$ are shown in Fig.~\ref{fig1}. Figure~\ref{fig1}(a) shows that the valence quark, gluon, and sea quark distributions for the pion in comparison with the experimental data of the E615~\cite{Conway:1989fs} and E615-M~\cite{Aicher:2010cb}. The valence quark distributions for the pion at $Q^2 =$ 5 GeV$^2$ have an excellent agreement with the data of Ref.~\cite{Conway:1989fs} for $m_u/m_d = 0.0$ (CS). However, the difference among the valence quark distributions for various values of $m_u/m_d$ is difficult to see due to the lines in the figure overlap. Similarly, it also occurs for the gluon and sea quark distributions for the pion.

To clearly see the CSB effects on gluon distributions for the pion. One shows that the ratios of the gluon distributions for the pion with different values of $m_u/m_d$ to those with $m_u/m_d =1.0$ (normalized) at $Q^2 =$ 5 GeV$^2$ as depicted in Fig.~\ref{fig1}(b). Figure~\ref{fig1}(b) clearly shows that the ratios of $g_{\pi^+} (x,m_u/m_d)/g_{\pi^+} (x,1.0)$ = 1.0 for $m_u/m_d = 1.0$ (solid line) which indicates the chiral symmetry gluon distributions for the pion. For $m_u/m_d = 0.7$ (dotted dashed line), $g_{\pi^+} (x,m_u/m_d)/g_{\pi^+} (x,1.0) \lesssim$ 1.0 and it slowly decreases as the longitudinal momentum of $x$ increases. Moreover, for $m_u/m_d = 0.5$ (dotted-dotted dashed line), the $g_{\pi^+} (x,m_u/m_d)/g_{\pi^+} (x,1.0) < g_{\pi^+} (x,0.7)/g_{\pi^+} (x,1.0)$ at $x \simeq 1$ and it rapidly decreases as $x$ increases. For $m_u/m_d = 0.3$ (dashed dashed line), the $g_{\pi^+} (x,m_u/m_d)/g_{\pi^+} (x,1.0)$ becomes larger than those for $m_u/m_d = 0.7$ and $m_u/m_d = 0.5$. The $g_{\pi^+} (x,m_u/m_d)/g_{\pi^+} (x,1.0)$ for $m_u/m_d = 0.0$ has the largest CSB effect size among other values of $m_u/m_d$ and it also decreases as the $x$ increases. As the results, one finds that the gluon distributions for the pion decrease as the CSB effects increase and it is more pronounced in the higher-$x$. The size of the CSB effects on gluon DFs in the pion for these respective different values of $m_u/m_d = [1.0, 0.7, 0. 5, 0.3, 0.0]$ at $x \simeq 1$ are approximately suppressed by [0.0, 1.0, 1.3, 2.6, 5.5]\% level, respectively, in comparison with those for $m_u/m_d = 1.0$.

Results for the sea $\bar{u}$-quark distributions for the pion with various values of $m_u/m_d$ normalizing with that for the pion with $m_u/m_d = 1.0$ at a scale of $Q^2 =$ 5 GeV$^2$ are shown in Fig.~\ref{fig1}(c). It clearly shows that the ratios of $\bar{u}_{\pi^+} (x, m_u/m_d)/\bar{u}_{\pi^+} (x, 1.0) < 1.0$ for $m_u/m_d = 0.7$. It slowly suppresses as $x$ increases up to $x \simeq 0.6$ and then it slowly increases again up to $x \approx 1.0$. Starting from $x \simeq 0.6$ up to $x \approx 1$, the size of the CSB effects on sea $\bar{u}$-quark distributions for $m_u/m_d = 0.7$ is approximately about 1.0\%. For $m_u/m_d = 0.5$, the suppression of ratios of $\bar{u}_{\pi^+} (x, m_u/m_d)/\bar{u}_{\pi^+} (x, 1.0)$ is larger than that for $m_u/m_d = 0.7$. The size of the CSB effect for $m_u/m_d = 0.5$ is about 2.0\% at around $x \simeq 0.6$ up to $x \approx 1$. The size of the CSB effect with $m_u/m_d = 0.3$ is expected about 2.5\% at $x \simeq 0.6$ up to $x \approx 1$. The largest CSB effect is given by $m_u/m_d = 0.0$ which is about 5.5\% at around $x \simeq 0.6$ up to $x \approx 1$.
%
%
%

In Fig.~\ref{fig1}(d), it shows that, at $Q^2 =$ 5 GeV$^2$, the sea $d$-quark distributions for the pion with various values of $m_u/m_d$ normalizing with that for the pion with $m_u/m_d =1.0$. The level of the CSB effect size on sea $d$-quark distribution for the pion with $m_u/m_d = 0.7$ is estimated at about 1.0\% at around $x \simeq 0.6$ up to $x \approx 1$. For $m_u/m_d = 0.5$, the size of the CSB effect on sea $d$-quark distributions for the pion is approximately about 2.0\% at around $x \simeq 0.6$ up to $x \approx 1$. At around $x \simeq 0.6$ up to $x \approx 1$, for $m_u/m_d = 0.3$, the size of the CSB effect on sea $d$-quark distributions for the pion is approximately about 3.0\%. As for the sea $\bar{u}$-quark case, the largest size of the CSB effect on sea $d$-quark distribution for the pion is given by $m_u/m_d =0$, where it is estimated by 6.0\% at around $x \simeq 0.6$ up to $x \approx 1$.
%
%
%
%

Similarly, as for the pion case, results for the valence quark, gluon, and sea quark distributions for the kaon, as well as the ratios of the valence quark, gluon, and sea quark distributions with various values of $m_u/m_d$ to those with $m_u/m_d =1.0$ for the kaon at $Q^2 =$ 5 GeV$^2$, is depicted in Fig.~\ref{fig2}. Figure~\ref{fig2}(a) shows the valence quark, gluon, and sea quark distributions for the kaon after evolving \textit{via} NLO DGLAP QCD evolution at scale $Q^2 =$ 5 GeV$^2$. Unfortunately, no experimental data are available for valence quark, gluon, and sea quark DFs of the kaon, at the moment, to compare. Therefore, the new experimental data for the kaon are really needed to measure in future experiments to verify these prediction results and other theoretical results. In addition, as for the pion case, it is also rather difficult to clearly see the CSB effect from the distributions directly. So, one can do is that to evaluate the ratios of $g_{K^+} (x, m_u/m_d)/g_{K^+} (x,1.0)$ for the partonic contents for the kaon as shown in Fig.~\ref{fig2}(b).

Figure~\ref{fig2}(b) shows the ratios of $g_{K^+} (x, m_u/m_d)/g_{K^+} (x,1.0)$ for kaon with various values of $m_u/m_d$ at a scale of $Q^2 =$ 5 GeV$^2$. At $x \approx 1$, the size of the CSB effect with $m_u/m_d = 0.7$ for the kaon is approximately about 0.1\% in comparison with that for $m_u/m_d = 1.0$. For $m_u/m_d = 0.5$, the size of the CSB effect is larger than that for $m_u/m_d = 0.7$, which is about 0.3\% at $x \approx 1$. For $m_u/m_d = 0.3$, at $x \approx 1$, the CSB effect size on the ratios of $g_{K^+} (x, m_u/m_d)/g_{K^+} (x,1.0)$ is approximately about 0.5\%. The size of the CSB effect is more pronounced for $m_u/m_d = 0.0$ which is approximately about 1.0\% at $x \approx 1$.

In Fig.~\ref{fig2}(c), it shows that, at a scale of $Q^2 =$ 5 GeV$^2$, the sea $\bar{u}-$quark distributions for the kaon with various values of $m_u/m_d$, normalizing with that for the kaon with $m_u/m_d = 1.0$. One finds that, at around $x \simeq 0.35$, the size of the CSB effect with $m_u/m_d = 0.7$ on sea $\bar{u}-$quark distributions for the kaon is approximately about 0.3\%. At around $x \simeq 0.35$, with $m_u/m_d = 0.5$, the CSB effect size on sea $\bar{u}-$quark distributions for the kaon is about 0.5\%. In addition, the size of the CSB effect with $m_u/m_d = 0.3$ at around $x \simeq 0.35$ is about 0.8\%. For $m_u/m_d = 0.0$, the CSB effect size on sea $\bar{u}-$quark distribution for the kaon at $x \simeq 0.35$ is about 2.0\% in comparison with that $m_u/m_d = 1.0$.

Results for the sea $s-$quark distributions for the kaon with various values of $m_u/m_d$, normalizing with that for the kaon with $m_u/m_d = 1.0$, at the scale of $Q^2 =$ 5 GeV$^2$ are depicted in Fig.~\ref{fig2}(d). One finds that the size of the CSB effect on ratios of $s_{K^+} (x,m_u/m_d)/s_{K^+} (x,1.0)$ with $m_u/m_d = 0.7$ at around $x \simeq 0.65$ is estimated about 0.8\%. Also, for $m_u/m_d = 0.5$, the size of the CSB effect on ratios of $s_{K^+} (x,m_u/m_d)/s_{K^+} (x,1.0)$ at around $x \simeq 0.65$ is found to be about 2.0\%. At around $x \simeq 0.65$, the size of the CSB effect on ratios of $s_{K^+} (x,m_u/m_d)/s_{K^+} (x,1.0)$ for $m_u/m_d = 0.3$ is approximated about 3.0\%. Moreover, the size of the CSB effect on ratios of $s_{K^+} (x,m_u/m_d)/s_{K^+} (x,1.0)$ for $m_u/m_d = 0.0$ at around $x \simeq 0.65$ is estimated about 6.0\%.
%
%

After investigating the size of the CSB effects on gluon and sea quark distributions for the pion and kaon, it now turns to observe the size of the CSB effects on quark distributions for the pion and kaon as depicted in Fig.~\ref{fig3}. In Fig.~\ref{fig3}(a), it is shown that the ratios of the up and anti down quark distributions for the pion with various values of $m_u/m_d$ at the scale of $Q^2 =$ 5 GeV$^2$. One finds that the size of the CSB effect for $m_u/m_d = 0.7$ at around $x \approx 1$ is estimated to decrease about 3.0\% in comparison with those with $m_u/md = 1.0$. With $m_u/m_d = 0.5$, the CSB effect on ratios of the up and anti down quark distributions for the pion is estimated to decrease about 4.0\% at $x \approx 1$ in comparison with those with the ratios of $m_u/m_d = 1.0$. The CSB effect size on ratios of the up and anti down quark distributions for the pion with $m_u/m_d = 0.3$ is suppressed about 6.0\% at $x \approx 1$. Among other ratios of $m_u/md$ values, the largest suppression of the CBS effect is given by $m_u/m_d = 0.0$, where the size of the CSB is estimated at about 10.0\%.

Figure~\ref{fig3}(b) shows the ratios of the up and antistrange quark distributions for the kaon with various values of $m_u/m_d$ at the scale of $Q^2 =$ 5 GeV$^2$. One finds that, for $m_u/m_d < 1.0$, the size of the CSB effects for the kaon is estimated to only a few percent level difference in comparison with those with $m_u/m_d = 1.0$. It can be clear to see the CSB effects on the individual quark distributions for the kaon are depicted in Figs.~\ref{fig3}(c) and ~\ref{fig3}(d). In Fig.~\ref{fig3}(c), it is shown that the ratios of the up quark distributions for the kaon with various values of $m_u/m_d$ to those with $m_u/m_d = 1.0$. One finds that the size of the CSB effect for $m_u/m_d = 0.7$ at $x \simeq 0.6$ is suppressed about 0.3\% in comparison with that for $m_u/m_d = 1.0$. For $m_u/m_d = 0.5$, the size of the CSB effect on the ratios of the $u$-quark distribution in the kaon to that for $m_u/m_d = 1.0$ is decreased by 0.6\% at $x \approx 0.6$. A remarkable result is given by $m_u/m_d = 0.3$, where the size of the CSB effect increases by 0.1\% at around $x \simeq 0.05$ in comparison with the current-quark mass ratios of $m_u/m_d = 1.0$ and it decreases by 0.9\% at around $x \simeq 0.6$. Following the result for $m_u/m_d = 0.3$, such similar behavior of the CSB effect can also be seen for $m_u/m_d = 0.0$, where the size of the CBS effect increases by 0.06\% at around $x \simeq 0.05$ and it then decreases by 2\% at $x \simeq 0.6$ in comparison with that for $m_u/m_d = 1.0$.

The result for the ratios of the antistrange quark distributions of the kaon for various values of $m_u/m_d$ to that for $m_u/m_d =1.0$ at scale $Q^2 =$ 5 GeV$^2$ is depicted in Fig.~\ref{fig3}(d). Figure~\ref{fig3}(d) shows the ratio of the antistrange quark distribution in the kaon decreases at $x \simeq 0.2$ and it then increases up to $x \approx 1$ which is in contrast with the up quark distribution in the kaon in Fig.~\ref{fig3}(c). One finds that the size of the CSB effect for $m_u/m_d = 0.7$ decreases by only a few percent at around $x \simeq 0.2$ and it then increases by 0.7\% at $x \approx 1$ in comparison with that for $m_u/m_d = 1.0$. For $m_u/m_d = 0.5$, the size of the CSB effect increases by 1.5\% in comparison with that for $m_u/m_d = 1.0$ at $x \approx 1$, whereas the CSB effect size, for $m_u/m_d = 0.3$, increases by 2.4\% compared with that for $m_u/m_d =1.0$ at $x \approx 1$. For $m_u/m_d = 0.0$ the size of the CSB effect in the quark distribution for the kaon, at $x \approx 1$, also increases by 4.8\% in comparison with that for $m_u/m_d = 1.0$\emdash which is the largest size of the CSB effect among other values of $m_u/m_d$.

To verify the prediction results of the present work, the comparisons between the gluon distributions for the pion at $Q^2 =$ 5 GeV$^2$ and $Q^2 =$ 16 GeV$^2$ with the prediction results of the lattice QCD simulation~\cite{Fan:2021bcr} and the JAM Monte-Carlo global fit QCD analysis~\cite{Barry:2018ort} are depicted in Fig.~\ref{fig4}(a). One finds that the gluon distributions for the pion at the scale of $Q^2 =$ 5 GeV$^2$ have an excellent agreement with the lattice QCD simulation~\cite{Fan:2021bcr} and the JAM Monte-Carlo global fit QCD analysis~\cite{Barry:2018ort} which is closer to the central values of the lattice QCD simulation at the intermediate regime of $x$. In addition, it also indicates that the gluon distributions for the pion at $Q^2 =$ 5 GeV$^2$ are better fitted than that for $Q^2 =$ 16 GeV$^2$. Again, unfortunately, there is no data available for the gluon and sea quark distributions for the kaon, at the moment. Therefore, the new data for the pion and kaon from the future experiments of the EIC~\cite{Arrington:2021biu,Aguilar:2019teb}, EicC~\cite{Anderle:2021wcy} as well as the COMPASS ++/AMBER new QCD facility at CERN-SPS~\cite{Adams:2018pwt} are expected to verify the prediction results of the present work and, of course, it will also be a very great opportunity for constraining other theoretical model predictions.

To deeper understanding of the size of the CSB effects on gluon distributions for the pion and kaon, one can calculate the ratios of the gluon distributions for the kaon to that for the pion with various values of $m_u/m_d$ at the scale of $Q^2 =$ 5 GeV$^2$ as shown in Figs.~\ref{fig4}(b). Figure~\ref{fig4}(b) shows the ratios of the gluon distribution for the kaon to that for the pion at a scale of $Q^2 =$ 5 GeV$^2$ increase as the size of the CSB effects increase. The ratios of the gluon distributions for the kaon to that for the pion with $m_u/m_d = [0.7, 0.5, 0.3, 0.0]$ increase a few percent levels, respectively, in comparison with those for the $m_u/m_d = 1.0$. Besides these interesting results, it also confirms that, as for the CS case, the gluon DFs for the pion are larger than those for the kaon with various values of the $m_u/m_d$ which is indicated by the decreasing of the ratios of $g_K (x)/g_\pi (x)$ as the longitudinal momentum of $x$ increases which is consistent with the prediction results in Refs.~\cite{Hutauruk:2019jja} and~\cite{Chen:2016sno}.

\section{Summary} \label{summary}
%

To summarize, in this present work, I have investigated, for the first time, the charge symmetry-breaking effects that arise from the $u-$and $d-$quark mass difference on gluon and sea quark DFs for both the pion and kaon using the Nambu--Jona-Lasinio model which is a chiral quark effective theory of QCD with the help of the PTR scheme, simulating color confinement of QCD. In the NJL model, there are no gluons and sea quarks dynamics in the initial model scale of $Q_0^2$ because they are already integrated out from the NJL Lagrangian and absorbed into the local four-fermion coupling constant of $G_\pi$. Therefore, the gluon and sea quark distribution functions for both the pion and kaon are purely and dynamically generated from the NLO DGLAP QCD evolution. Finally, results for the gluon and sea quark distribution functions for both the pion and kaon are compared with the prediction results from the recent lattice QCD simulation~\cite{Fan:2021bcr} and JAM global fit QCD analysis~\cite{Barry:2018ort} to verify the findings of this present work.

I first evaluated the valence quark, gluon, and sea quark distribution functions for the pion and kaon at a higher factorization scale of $Q^2 =$ 5 GeV$^2$ \textit{via} the NLO QCD evolution of DGLAP. One finds that the result for the gluon distributions for the pion is in good agreement with the prediction results obtained from the recent lattice QCD simulation~\cite{Fan:2021bcr} and JAM global fit QCD analysis~\cite{Barry:2018ort}. Also, the result for the valence quark distributions for the pion at $Q^2 =$ 5 GeV$^2$ has an excellent agreement with the old E615 data~\cite{Conway:1989fs}.

The size of the CSB effects on gluon distributions for the pion with the realistic $m_u/m_d = 0.5$ at the scale of $Q^2 =$ 5 GeV$^2$ is estimated by 1.3\% in comparison with those with $m_u/m_d = 1.0$ (chiral symmetry) at $x \simeq 1$. Also, it generally decreases as the longitudinal momentum of $x$ increases. In addition, for the sea $\bar{u}-$quark distributions for the pion with the realistic $m_u/m_d = 0.5$ at a scale of $Q^2 =$ 5 GeV$^2$ is estimated by 2.0\% at around $x \simeq 0.6$ up to $x \simeq 1$. For the sea $d-$quark distributions for the pion with the realistic $m_u/m_d =0.5$ at $Q^2 =$ 5 GeV$^2$, the size of the CSB is approximately about 2.0\% at around $x \simeq 0.6$ up to $x \simeq 1$.

Next, the size of the CSB effects on gluon and sea quark DFs for the kaon is obtained. One finds that the size of the CSB affects the ratio of the gluon distributions for the kaon with the realistic $m_u/m_d = 0.5$ with that with $m_u/m_d = 1.0$ is about 0.3\% at $x \simeq 1$. The sea $\bar{u}-$quark distributions for the kaon with the realistic $m_u/m_d = 0.5$ is given by 0.5\% at $x \simeq 0.35$. The result for the size of the CSB effects on the ratios of the sea $s-$quark distribution for the kaon with $m_u/m_d =0.5$ with that with $m_u/m_d = 1.0$ is approximately about 2\% at $x \simeq 0.65$.

Moreover, remarkable results have been found that the size of the CSB effect on gluon distribution functions for the pion is larger than that for the kaon. Similarly, as for the charge symmetry case, it also gives that the gluon distribution functions for the pion are larger than that for the kaon which is consistent with the prediction results in Refs.~\cite{Hutauruk:2019jja} and~\cite{Chen:2016sno}.

Overall, one can conclude that the prediction results for the CSB effect are rather small and it could be very challenging for future experiments. Therefore, the finding results of the present work provide crucial information for testing the predictions on the expected size of the CSB effects on gluon and sea quark structure functions for the kaon and pion in the experiments which have been outlined in Ref.~\cite{Londergan:1994gr} as well as the EIC~\cite{Arrington:2021biu,Aguilar:2019teb}, EicC~\cite{Anderle:2021wcy} as well as the COMPASS ++/AMBER new QCD facility at CERN-SPS~\cite{Adams:2018pwt}. Experimentally, the CSB or CSV effects in the PDFs will be planned to indirectly measure in Solenoidal Large Intensity Device (SoLiD) at JLab12 upgrade and beyond through the parity-violating deep inelastic scattering process~\cite{zein2022}.

Also, the results of this work would serve as a very useful guidance and tool to investigate the size of the CSB effect on gluon and sea quark distribution functions for the pion and kaon using the lattice QCD simulations which is the first principle approach and other realistic and sophisticated theoretical approaches.

Further investigations on gluon distribution functions for the kaon and pion and other types of mesons as well as for the nucleon which has a more complicated structure than meson are really needed for a deeper understanding of the role of the CSB effect in hadron structure functions. A study on gluon content for the nucleon including the CSB effect is ongoing and the result will be expected to appear somewhere in the near future.

\begin{acknowledgments}
This work was partially supported by the National Research Foundation of Korea (NRF) funded by the Korean government (MSIT) Grants No.~2018R1A5A1025563 and No.~2022R1A2C1003964.
\end{acknowledgments}

\end{document}